\documentstyle[preprint,aps,prb]{revtex}
\tightenlines

\begin{document}
\draft
\preprint{HEP/123-qed}
\title{Mechanism of carrier-induced ferromagnetism \\
in magnetic semiconductors }
\author{Masao Takahashi}
\address{
Kanagawa Institute of Technology \\
1030 Shimo-Ogino, Atsugi-shi 243-0292, Japan \\ 
}
\author{Kenn Kubo}
\address{Department of Physics,  
Aoyama Gakuin University \\
Setagaya,  
Tokyo 157-8572, Japan
}
\date{\today}
\maketitle
\begin{abstract}
Taking into account both random impurity distribution and thermal
fluctuations of localized spins, 
we have performed a model calculation for the carrier (hole) state in Ga$_{1-x}$Mn$_x$As
by using the coherent potential approximation (CPA).  
The result reveals that a {\it p}-hole in the band tail of
Ga$_{1-x}$Mn$_x$As is not like a free carrier
but is rather virtually bounded to impurity sites.
The carrier spin strongly 
 couples to the localized {\it d} spins  on Mn ions.
The hopping of the carrier among Mn sites
causes the ferromagnetic ordering of the localized spins
through the double-exchange mechanism.
The Curie temperature obtained by using conventional parameters agrees well with the experimental result. 
\end{abstract}
\vspace{0.5cm}
\narrowtext

%
 Although a considerable amount of experimental data has already been
accumulated, the origin of the ferromagnetism in III-V-based diluted
magnetic semiconductors (DMS's) has still not been clarified theoretically.
Since the magnetic interaction between Mn ions has been shown to be
antiferromagnetic in $n$-type (In,Mn)As or intentionally compensated (Ga,Mn)As,
the ferromagnetic interaction in III-V-based DMS's is most likely
 hole-induced.\cite{rev1} 
The ferromagnetism may be caused by the effective exchange interaction
between the delocalized carrier (or the hole originating from the shallow acceptor)
 and the localized magnetic moment ({\it d} spin) on the Mn ion.
Matsukura {\it et al.} \cite{Matsukura} referred to the
Ruderman-Kittel-Kasuya-Yoshida (RKKY) interaction as the origin of the
above ferromagnetism.
However, the application of the RKKY interaction to this system, as pointed out by various authors, \cite{Sakai,Akai,Inoue} is somewhat questionable.
The RKKY formula is applicable when the exchange energy
 $IS\ (\equiv I \times S)$
is small compared to Fermi energy, 
but this is not the case here because of the low carrier
density.\cite{Sakai}  
Furthermore, the measurement of the magnetic circular dichroism (MCD)
spectrum of (Ga,Mn)As suggests that the holes play more active role in
mediating the ferromagnetic exchange  than traditional RKKY
systems.\cite{Beschoten} 
To date, in regard to the origin of the above ferromagnetism, 
the double-exchange mechanism,\cite{Akai}  
the double-resonance mechanism,\cite{Inoue} the 
Zener-model description,\cite{Dietl} 
and the self-consistent theory based on the CPA and mean-field theory
\cite{Kayanuma} have been proposed.

%
In this letter, we apply the dynamical coherent potential approximation 
(dynamical CPA) \cite{Kubo,Edward,takaCPA} to 
 the carriers in Ga$_{1-x}$Mn$_x$As described by a 
model 
Hamiltonian $H$;
\cite{takaDMS1,takaDMS2}
\begin{eqnarray}
 H &=& \sum_{m,n,\mu} \varepsilon_{mn} a^{\dag}_{m\mu} a_{n\mu}  +
\sum_{m} U_m,
\end{eqnarray}
 where $a^{\dag}_{m\mu} \ (a_{m\mu})$ creates (annihilates) a carrier (or $p$-hole) with spin $\mu$ at 
the $m$-th site.
  We treat Ga$_{1-x}$Mn$_x$As  as an alloy
in which a mole fraction $x$ of Ga ions (denoted by symbol $A$) is
replaced at random with Mn ions (denoted by symbol $M$).
 Hence $U_m$ is
 $E_A  \sum_{\mu}  a^{\dag}_{m\mu} a_{m\mu}$ or 
$E_M \sum_{\mu}  a^{\dag}_{m\mu} a_{m\mu}
-I \sum_{\mu\nu} a^{\dag}_{m\mu} \mbox{\boldmath $\sigma$}_{\mu\nu} \cdot
{\bf S}_m a_{m\nu}$
depending on whether the site $m$ is occupied by a Ga  
or a Mn ion.
Here, $E_A$ and $E_M$ denote the spin-independent potentials for the Ga
 and the Mn ion, respectively;
$-I \mbox{\boldmath $\sigma$} \cdot {\bf S}_m$ represents the {\it p-d}
exchange interaction between the carrier and the localized 
({\it d-}) spin ${\bf S}_m$ of Mn at the site $m$.
Hereafter, we set $E_A \equiv 0$ as the origin of the energy. 
A carrier moving in a DMS is subject to disordered potentials
which arise not only from substitutional disorder but also from thermal
fluctuations of the localized 
({\it d-})  spins.
In the dynamical CPA,\cite{takaCPA}  the disordered potentials are taken 
into account in terms of the spin-dependent coherent potentials,
$\Sigma_{\uparrow}$ and $\Sigma_{\downarrow}$,  
which are determined such that 
the effective scattering of a carrier at an arbitrarily chosen
site embedded in the effective medium vanishes on average.
The condition (the dynamical CPA condition) is given by
\begin{mathletters}
\label{CPA}
\begin{eqnarray}
(1-x) t^A_{\uparrow \uparrow} + x <t^M_{\uparrow \uparrow}>_{\rm av} &=&
0 \ ,  \label{CPA:1}  \\
(1-x) t^A_{\downarrow \downarrow} + x <t^M_{\downarrow \downarrow}>_{\rm
av} &=& 0  \ . \label{CPA:2}
\end{eqnarray}
\end{mathletters}
%
The multiple scattering of an $\uparrow (\downarrow)$-spin  carrier  by an
$A$ and an $M$ ion 
embedded in the effective medium is  represented by
the {\it t} matrix element 
$t^A_{\uparrow \uparrow}\  
(t^A_{\downarrow \downarrow})$  and $t^M_{\uparrow \uparrow}\  
(t^M_{\downarrow \downarrow}) $, respectively.
 Note that in the above expression for $t^M_{\uparrow \uparrow} 
\ (t^M_{\downarrow \downarrow}$), the spin flip processes are  properly
taken into account. \cite{takaCPA}
The thermal average over the fluctuations of the localized 
 spins is taken by assuming that they are subject to an
effective field $h$, and  denoted by $<\ >_{\rm av}$.
%
%
In this work we assume  the localized 
 spins to be classical.
Assuming a semicircular  density of states(DOS)
with the half-bandwidth $\Delta$ for the host valence band, we solved 
 Eqs (\ref{CPA:1}) and (\ref{CPA:2}) simultaneously,  
and obtained the DOS  per site $D_{\uparrow}(\omega)\ (D_{\downarrow}(\omega))$ of the carrier  with
 $\uparrow \ (\downarrow)$ -spin.

%

We assume $\Delta = 2$eV, $IS/\Delta = -0.4$ and $E_M/\Delta = -0.3$
as parameters for Ga$_{1-x}$Mn$_x$As;
$E_M$  ($< 0 $) is the attractive potential acting  
on a hole  due to a Mn$^{2+}$ ion
(see later discussion).
 In the diluted impurity limit ($x \to$ 0),  
the present model gives an impurity (acceptor) level at the energy of
$E_a = -1.057\Delta$. 
 With increase in Mn concentration an impurity band forms,  and 
for $x >0.02$ it merges into the host band which
originates 
from the valence band of the host semiconductor.
This result is consistent  with  the observed insulator- metal transition in Ga$_{1-x}$Mn$_x$As at  $x \sim 0.03$.\cite{Oiwa} 
 The  result of  the  carrier DOS for $x$ =0.05
is shown in Fig.\ 1(a)
for various values  of magnetization $<S_z>_{\rm av}/S$.
 Note that the system with $x=0.05$ is metallic.
The band tail arising from the magnetic impurity band 
is strongly affected by magnetization. 
 With increase of the magnetization the lower edge of the DOS is 
slightly shifted  to the lower energy and at the same time 
$D_{\uparrow}(\omega)$ 
is strongly suppressed in the band tail.  We show the optical carrier spin polarization 
defined by
%
%
%
$P(\omega) = (D_{\uparrow}(\omega)-D_{\downarrow}(\omega))
/(D_{\uparrow}(\omega)+D_{\downarrow}(\omega))$
 in Fig.1(b).
 In order to examine the exchange coupling of carriers with localized spins,  we  calculated  
 the relative local DOS at Mn sites,
$R(\omega) = (D^{\rm M}_{\uparrow}(\omega)+D^{\rm M}_{\downarrow}(\omega))
/(D_{\uparrow}(\omega)+D_{\downarrow}(\omega))$, and depicted it  in  Fig.1(c).
The result shows that a  carrier at  the band tail stays 
mainly at  Mn sites $(R(\omega) \sim 0.6)$ 
in spite of small  $x$.
It should be noted that $R(\omega)$ is almost independent 
of $<S_z>_{\rm av}$.
The result suggests 
that the change in $P(\omega)$ is mainly 
ascribed to the change at the magnetic ion sites.

In order to examine how strongly the carrier spin couples to the localized spins 
we calculate the carrier spin polarization 
$P(n) \equiv (n_{\downarrow}-n_{\uparrow})/(n_{\downarrow}+n_{\uparrow})$
as a function of the total carrier density $n$.
Throughout this study, we assume that the carriers are degenerate.
%
In Fig.2, we show the results of $P(n)$ for
 $x=0.01$ and $x=0.05$.
When $x=0.01$,
the impurity band is separated from the host band, so that
carriers stay mainly at Mn sites as long as $n \lesssim x$. 
 The carrier spin at a Mn site is expected to follow faithfully 
 the fluctuations of the localized spin. 
This leads to the result that 
 $P(n) \approx <S_z>_{\rm av}/S$ for $n \lesssim x \ (=0.01)$, 
as shown in Fig.2(a).
When $x=0.05$,
the magnetic impurity band partly merges into the host band
and  
the Fermi level ($\varepsilon_F$) 
reaches the bottom of the host band
at the carrier density $n_1\ \sim 0.035$.
Accordingly,  $P(n) \approx <S_z>_{\rm av}/S$ for $n \lesssim n_1$ 
as shown in Fig.2(b).
 When $n \gtrsim n_1$,  carriers with energy higher than  $ -\Delta$
seeps out of Mn site to itinerate among nonmagnetic sites.

 We calculate 
the total energy 
  as a function of $<S_z>_{\rm av}$ as
\begin{eqnarray}
E(<S_z>_{\rm av}) &=& \int_{-\infty}^{\varepsilon_F}  \omega
[D_{\uparrow}(\omega)+D_{\downarrow}(\omega)] 
d \omega \ .
\end{eqnarray}
Note that  $E(<S_z>_{\rm av})$ is the sum of 
the kinetic and  exchange energy.
 We show in Fig.3(a) 
the energy difference  between the ferromagnetic  and
the paramagnetic state
 as a function of $n$.
Figure 3(a) reveals that the energy gain 
is approximately proportional to $\left( \frac{<S_z>_{\rm av}}{S} \right)^2$
 at a fixed $n$.

The free energy of the present model per site is given as
\begin{eqnarray}
F(<S_z>_{\rm av}) &=& E(<S_z>_{\rm av})
 -  T {\cal S} ,
\end{eqnarray}
where the entropy term due to the localized spins is given by
\begin{eqnarray}
 {\cal S} & = & x k_B \log \sum_{S_z = -S}^S {\rm exp} \left( \frac{h
S_z}{k_B T}\right)  - x \frac{h}{T} <S_z>_{\rm av} \ .
\end{eqnarray}
The effective field $h$ 
is determined so as to minimize $F(<S_z>_{\rm av})$ through the condition $\frac{d}{d h} F(<S_z>_{\rm av}) =0$.
 If we can expand   $F(<S_z>_{\rm av})$
in terms of $(<S_z>_{\rm av})^2$,  $T_c$ is determined as the temperature 
where the coefficient of $(<S_z>_{\rm av})^2$ vanishes. 
While calculations for small magnetization accurate enough to estimate the coefficient is difficult,
 we observed from Fig.3(a) that $E(<S_z>_{\rm av})-E(0)]$ is approximately proportional
 to $(<S_z>_{\rm av})^2$ up to full polarization. 
Therefore  
we fitted the data of $[E(<S_z>_{\rm av})-E(0)]/\Delta$ with
 $(<S_z>_{\rm av}/S)^2=$ 0.4 and 1.0 to 
the expansion 
$E(<S_z>_{\rm av})-E(0)]/\Delta = -a(<S_z>_{\rm av}/S)^2
+b(<S_z>_{\rm av}/S)^4$ and estimated  $a$ and $b$
 in actual calculation. 
Then $T_c$ is given  by $k_B T_c = 2a/3x$.
In Fig.3(b), the result of $T_c$  is depicted as a function of $n$
  for $x=0.05$. 
It has been reported that
 ferromagnetism with Curie temperature $T_c=110 $K is
realized in 
Ga$_{1-x}$Mn$_x$As with $x=0.053$ when $n$ is 30\% of the nominal
concentration of Mn.
\cite{rev1}
  The present result is in excellent agreement with the experimental result.

In order to obtain further insight into the mechanism of the carrier-induced
ferromagnetism in DMS's,  
we consider the case that the exchange interaction
is so strong that the carrier spins keep almost antiparallel  
 to the fluctuating localized spins. 
 We show the result for $IS/\Delta = -1.0$ and $E_M/\Delta=0$
 in Fig. 4 for illustration.
The magnetic impurity band is formed about the impurity level
(marked 'A') 
 and is separated from the host band.
The total number of states in the impurity band per site
is equal to $x$ 
irrespective to the value of $<S_z>_{\rm av}/S$.
When $<S_z>_{\rm av}=S$, 
all states in the impurity band 
are down-spin state,
while when $<S_z>_{\rm av}=0$ 
the impurity band is composed of the same number ($x/2$)
of up- and down-spin states. 
The ferromagnetic DOS extends over a 
wider energy range about the impurity level
 than does the paramagnetic DOS.
Consequently the ferromagnetic state has lower energy
 than the paramagnetic state when $n$ is small.
 This result implies that the 
double-exchange mechanism \cite{Zener}
works in the magnetic impurity band   
to realize ferromagnetism.
  The energy gain increases first with the increase of $n$ and 
 reaches a maximum value at
$n \sim x/2$. 
Then it decreases gradually and vanishes finally  at
$n \sim x$ as shown in Fig. 3(b). 
 The same mechanism  is expected to be operative in the case of 
$IS =-0.4 \Delta $ and $E_{\rm M} = -0.3\Delta$ 
although the impurity band is not separated in this case.
%
%
%
 Therefore we conclude that the ferromagnetism 
in Ga$_{1-x}$Mn$_x$As with $x=0.05$
is caused by the double-exchange mechanism in the band tail.

Here, we discuss the parameters and the model 
 used in the present work.
 Since the width of the valence band is estimated to be about 4 eV
by  band calculations,
\cite{Shirai,Park} 
we have taken $\Delta =2$ eV.
There are controversial discussions on the magnitude (even on the sign)  of the exchange
constant  in III-V-based DMS's. \cite{rev1}
Therefore, 
we have assumed that $IS/\Delta=-0.4$,
which is a conventional value in II-VI-based DMS's. \cite{takaDMS1}
Note that $|\frac{E_M + IS}{\Delta}|>0.5$
is the necessary condition for the existence of an impurity level.
Accordingly, the attractive potential $E_M<0$ is needed
to reproduce an impurity level, which is confirmed by  core-level photoemission experiment. \cite{Okabayashi}
 Since the impurity level ${E_a}$ is given by 
$\frac{E_a}{\Delta} =
\left(\frac{E_M + IS}{\Delta}\right)+
\frac{1}{4}\left(\frac{\Delta}{E_M + IS}\right)$ 
in the present model, 
 we introduced the attractive potential $E_M= -0.3\Delta$,
 which 
reproduces experimental value of the acceptor energy  
 0.113eV $(=0.057 \Delta)$.
 \cite{Linnarsson}
%
We emphasize that 
the attractive potential strongly assists the occurrence
of the ferromagnetism.
If we take $E_M/\Delta =0$ together with $IS/\Delta=-0.4$,
which corresponds to Mn doped II-VI DMS's,
 $T_c$ is calculated to be at highest 40 K for $x=0.05$.
This is because the carrier in the II-VI-based DMS's
spreads over many nonmagnetic sites,
which reduces the effective exchange coupling. 
The result explains well why III-V-based DMS's such as (Ga,Mn)As exhibit
appreciable ferromagnetism while II-VI-DMS's such as (Cd,Mn)Te and
(Zn,Mn)Te exhibit only negligible ferromagnetism.
To the contrary, if we take $E_M/\Delta = -0.6$ together with $IS/\Delta=-0.4$,
the magnetic impurity band separates from
the host band, and $ T_c$ increases to 140 K for $x=0.05$.
In the latter case, the carrier is semi-localized to a magnetic impurity
site by the attractive potentials so that
the double-exchange mechanism works efficiently.

In the present model, the carrier transfer energy 
$(\varepsilon_{mn})$ between the Mn site and Ga site is assumed to be the same as that between Ga sites. 
 That is, holes in (Ga,Mn)As are
assumed to have mainly the As 4$p$ character
which is originated from the valence-band maximum of the host GaAs, rather than Mn 3$d$ character 
(although we actually treated only carriers on cation sites).
Therefore, the present model is in contrast to the double-exchange
model based on the Mn 3$d$ band. \cite{Akai} 
 Results of recent studies using the angle-resolved photoemission
spectroscopy \cite{Oka3} and the band-structure calculation using the
local-density-approximation with Coulomb interaction (the LDA+U method)
\cite{Park} strongly support our assumption. 

 In summary, we have shown that the  carriers in the band
tail of Ga$_{1-x}$Mn$_x$As have so large
 local-carrier-density on magnetic ion sites that the
carrier spins well couples the localized spins fluctuating
on the Mn ions.
The double-exchange mechanism in the magnetic impurity band causes the
ferromagnetism in the DMS.
The attractive local potentials at impurity sites
 strongly assist the occurrence of the ferromagnetism in III-V-based DMS's.

\acknowledgments
The authors would like to thank Professor Y. Kayanuma for informing them of the work of his group before publication.
One of the authors (M. T.) is grateful to Professor W. Nolting for his
instructive comments on the carrier-induced ferromagnetism.
 K. K. was  partially  supported by Center for Science and Engineering Research, Research Institute of Aoyama Gakuin University.

\begin{figure}
\caption{(a) The Carrier (hole) density of states for 
$x$=0.05 are depicted for 
  various values of magnetization. 
The solid and  dotted lines represent  
$D_{\downarrow}(\omega)$ and 
$D_{\uparrow}(\omega)$, respectively. 
The arrow indicates the Fermi level for $ n=x\ (=0.05) $,  which is
 about  $-0.95\Delta$ and is  almost independent  of 
$<S_z>_{\rm av}$.
 (b) The optical carrier spin polarization defined by 
$P(\omega) = [D_{\downarrow}(\omega)-D_{\uparrow}(\omega)] /[D_{\downarrow}(\omega)+D_{\uparrow}(\omega)]$.
The impurity level $(\omega/\Delta =-1.057)$
is also indicated by a dot (symbol 'A').
(c) The ratio of the local DOS at Mn site to the total DOS, $R(\omega) = [D^M_{\uparrow}(\omega)+D^M_{\downarrow}(\omega)]
/[D_{\uparrow}(\omega)+D_{\downarrow}(\omega)$], for
$<S_z>_{\rm av}/S$ = 0.0, 0.5, and 1.0. 
} 
\label{autonum}
\end{figure}

\begin{figure}
\caption{The carrier spin polarization defined by 
$P(n) = (n_{\downarrow}-n_{\uparrow})/(n_{\downarrow}+n_{\uparrow})$
for various degrees of magnetization for $x=0.01$ (a) and  $x=0.05$ (b).
The arrows indicate  $n=x$ and $n=n_1$; $n_1$ is the 
carrier density where the  Fermi level ($\varepsilon_F$) reaches the bottom of the
host band.
}
\label{autonum}
\end{figure}

\begin{figure}
\caption{
(a) $[E(<S_z>_{\rm av})-E(0)]/\Delta$ is depicted for $x=0.05$ as a function of the carrier density $n$,
 for various values of
$(<S_z>_{\rm av}/S)^2$.
(b) The Curie temperature $T_c$ calculated for $x=0.05$
as a function of $n$.
}
\label{autonum}
\end{figure}

\begin{figure}
\caption{
The DOS of the magnetic impurity band in the case of $IS/\Delta = -1.0$ 
and $E_M /\Delta = 0$.
The thick, thin and dotted lines show the cases of 
 $<S_z>_{\rm av}/S=1.0$, 
$0.5$ and 0.0, respectively.  
The magnetic impurity level is dotted with symbol 'A'.}
\label{autonum}
\end{figure}

\end{document}